\documentclass[aps,prb,twocolumn,showpacs]{revtex4}
\usepackage{amsmath}
\usepackage{graphicx,epsfig,psfrag}
\usepackage{amssymb}
\usepackage{bm}

\newcommand{\be}{\begin{equation}}
\newcommand{\ee}{\end{equation}}
\newcommand{\bea}{\begin{eqnarray}}
\newcommand{\eea}{\end{eqnarray}}

\newcommand{\w}{\omega}

\newcommand{\s}{\sigma}

\renewcommand{\Re}{\mathrm{Re}\;}

\begin{document}

\title{Spin conductivity in almost integrable spin chains}
\author{Peter Jung}
\author{Achim Rosch}
\affiliation{Institute for Theoretical Physics, University of
Cologne, 50937 Cologne, Germany.}
\begin{abstract}
The spin conductivity in the integrable spin-$1/2$ XXZ-chain is known to be infinite at finite temperatures $T$ for anisotropies $-1<\Delta<1$. Perturbations, which break integrability, e.g.~a next-nearest neighbor coupling $J'$, render the conductivity finite. We construct numerically a non-local conserved operator $J_\|$ which is responsible for the finite spin Drude weight of the integrable model and calculate its decay rate for small $J'$. This allows us to obtain 
 a lower bound for the spin conductivity $\s_s \geq c(T)/ J'^{2} $, where $c(T)$ is finite for $J' \to 0$. We discuss the implication of our result for the general question how non-local conservation laws affect transport properties.
\end{abstract}
\date{\today}
\maketitle
\section{Introduction}   
The behavior of transport properties of integrable systems have attracted considerable interest in the recent past \cite{zotos:1997,naef.zotos:1998,fabricius.mccoy:1998,narozhny.millis.andrei:1998,zotos:1999,rosch.andrei:2000,klumper.sakai:2002,heidrich-meisner:2002,heidrich-meisner:2003,fujimoto.kawakami:2003,shimshoni:2003,prelovsek:2004,klumper:2005,heidrich-meisner:2005-b,jung.rosch:2006,jung.rosch:2007,sologubenko:2001,kinoshita:2006,hess:2007,sologubenko:2007}. 
In such models due to the presence of conservation laws the currents do not decay.\cite{zotos:1997}
As a consequence, the dc conductivity is infinite and characterized by a finite Drude weight $D$, 
$\Re \s(\w)=\pi D \delta(\w)+\s_\text{reg}(\w)$, where $\s_\text{reg}$ is the regular part of the conductivity.

In real systems those conservation laws are violated by perturbations
 which often can be considered to be small. In these situations, the conductivity 
becomes finite \cite{zotos:1997,rosch.andrei:2000,heidrich-meisner:2003,shimshoni:2003,jung.rosch:2006,jung.rosch:2007} but remains very large as long as the perturbations are small.

A both theoretically\cite{grabowski.mathieu:1995,naef.zotos:1998,fabricius.mccoy:1998,narozhny.millis.andrei:1998,zotos:1999,klumper.sakai:2002,heidrich-meisner:2002,fujimoto.kawakami:2003,heidrich-meisner:2005-b,jung.rosch:2006} and experimentally\cite{sologubenko:2001,sologubenko:2007,hess:2007} well studied example is the XXZ Heisenberg chain which is equivalent to a model of spinless Fermions with nearest neighbor interactions.

The XXZ Heisenberg chain is integrable and therefore an infinite number of constants of motion exists for this model. All eigenstates can be uniquely labeled by a complete set of commuting operators, $Q_n$ with $[Q_n,Q_m]=0$ for all $n,m$. The first two of these operators are given by the total magnetization $Q_1=\sum_k S^z_k$ and the Hamiltonian
\bea
H_0&=&Q_2=\sum_i h_i,\label{XXZ}\\
 h_i&=&J ( S_i^x S_{i+1}^x+S_i^y S_{i+1}^y+ \Delta S_i^z S_{i+1}^z)
\eea
All other $Q_n$ can be constructed by a simple recursive formula\cite{grabowski.mathieu:1995}, $Q_{n+1}= [B,Q_n]$ with the so-called boost operator $B=1/(2 i)\sum_j j h_j$. All these conservation laws have a property which is important for the following discussion: they 
are {\em local} operators in the sense that each $Q_n$ can be written in terms of  a local ``density'' $q_{n,i}$  at site $i$,
\bea
Q_n=\sum_i q_{n,i} \label{local},
\eea
where $q_{n,i}$ is local as it contains only spin operators $S^\alpha_j$ on  maximally $n$ adjacent sites,
$i\le j < i+n$. 

Besides the $Q_n$, there exists a huge number of other conservation laws $C_i$ which can in principle be constructed from the exact eigenstates $|j\rangle$, $[H_0,\, |j\rangle \langle j' |\,]=0$ for 
$E_j=E_{j'}$. In general, these operators are highly non-local objects in the sense that they cannot be written in the form (\ref{local}) for finite $n$.

Only local conservation laws are associated with a continuity equation $\partial_t q_{n,k}+ j_{n,k+1}-j_{n,k}=0$, where $j_{n,k}$ is the corresponding current density, and therefore only for local conservation laws a hydrodynamic description can be formulated.
A main motivation for the present work is the question, to what extent non-local
conservation laws are relevant in the sense that they lead to experimentally observable consequences in real materials. We therefore study the role of local and non-local conservation laws for transport in XXZ Heisenberg chains perturbed by weak next-nearest neighbor couplings $J'$.

For $J'=0$,   both the heat conductivity $\kappa$ and the spin conductivity $\sigma_s$ (or, equivalently, the electric conductivity in the Fermionic language) are infinite and have a finite Drude weight. However, there is a main conceptual difference between those two cases: the heat current $J_E$ is conserved (and actually given by $Q_3$ as defined above), while the spin current $J_s$ does not commute with the Hamiltonian. Nevertheless, the presence of a finite Drude weight implies that a certain fraction of the spin current does not decay in time: part of the spin current is `protected' by conservation laws. This has been formalized many years ago by Mazur\cite{mazur:1969} and later generalized by Suzuki\cite{suzuki:1971}. Suzuki showed that
the Drude weight can be expressed in terms of correlators of the current with the conservation laws $C_i$,
\bea
D_s=\frac{\beta}{N} \sum_i \frac{\langle J_s C_i\rangle^2}{\langle  C_i^2\rangle} , \label{suz}
\eea
where $\beta=1/T$, $N$ is the number of sites and the $C_i$ have been chosen such that $\langle  C_i C_j\rangle=0$ for $i \neq j$. Note that in Eq.~(\ref{suz}) the sum runs over a basis of {\em all} conservation laws, local and non-local, commuting and non-commuting.

Interestingly, it can be shown \cite{zotos:1997}  by simple symmetry arguments that the spin current is orthogonal to all known local conservation laws, $\langle J_s Q_n\rangle=0$ for all $n$. Therefore it seems that non-local conservation laws are responsible for the finite conductivity of the integrable model.
What will happen to the spin current when the system is weakly perturbed, e.g.~by a next-nearest neighbor coupling $J'$
with $J' \ll J$? For local conservation laws, e.g.~the heat current $Q_3$, the answer is known:\cite{jung.rosch:2006}
for small $J'$, $Q_3$ decays only slowly implying a large dc conductivity proportional to the 
life-time of $Q_3$. Our present goal is to investigate whether the spin-conductivity shows a similar behavior. 

An example which shows that perturbation theory for local and non-local quantities can be drastically different has been discussed in Ref.~[\onlinecite{anfuso:2007}]. In this paper it has been shown that an arbitrarily small inter-chain coupling can destroy a non-local order parameter (e.g.~the string order of a spin-1 Haldane chain) in a gapped system. Formally, the perturbations turn out to be proportional to the length of the system. In contrast, local order parameters are always robust against small perturbations for all gapped systems.

In principle one can try to investigate the transport properties for small $J'$  directly by calculating the spin-conductivity from an exact diagonalization of the XXZ chain in the presence of  finite $J'$. In such a calculation Heidrich-Meisner {\it et al.} \cite{heidrich-meisner:2003} were able to show that the spin  Drude weight vanishes in the thermodynamic limit, but a reliable determination of the resulting finite spin-conductivity is rather difficult even for large $J'$. Furthermore, finite size effects grow rapidly\cite{heidrich-meisner:2003} for small $J'$. 

In the following, we will therefore use a different approach based on a perturbation theory in $J'$. We construct numerically a non-local operator $J_\|$ which is conserved for $J'=0$ and responsible for the finite Drude weight of the unperturbed XXZ Heisenberg chain.
In a second step we derive a lower bound for the spin conductivity of the perturbed system using results of Ref.~[\onlinecite{jung.rosch:2007}] and show that the spin conductivity is proportional to $1/J'^2$.
Finally, we analyze to what extent  $J_\|$ is a non-local operator and discuss how the result can be interpreted. Appendix \ref{appA} investigates the role of special values of the anisotropies where the Heisenberg model possesses extra symmetries.
\section{Model} 
We consider the following Hamiltonian
\be
H=H_0+H_1 ,
\ee
where the XXZ Heisenberg chain $H_0$ has been defined in Eq.~(\ref{XXZ}) and
\be
H_1=J'\sum_k S_k^x S_{k+2}^x+S_k^y S_{k+2}^y+ \Delta S_k^z S_{k+2}^z 
\ee
describes the (small) next-nearest neighbor coupling.
For this model,  the spin current $J_s$ is given by
\be
J_s=\frac{i}{2} J \sum_k \left( S^+_k S^-_{k+1}-S^-_k S^+_{k+1} \right)+O(J')
\ee
and we have omitted terms linear in $J'$ as they give only subleading contributions to our final result.

For $J'=0$ and $-1<\Delta<1$, the Drude weight defined by 
\be
\Re \s_s(\w)=\pi D \delta(\w)+\s_{\rm reg}(\w)
\ee
is finite\cite{narozhny.millis.andrei:1998,zotos:1999,klumper:2005} at $T>0$ as discussed above. Equation (\ref{suz}) implies that 
the finite Drude weight is associated to constants of motion $C_i$ of $H_0$ with
$\langle C_i J_s\rangle \neq 0$ which we need to identify for our further analysis. 
More precisely, one can split the current operator into two pieces, 
\be
J_s=J_\|+J_\perp , \label{sep}
\ee
with
\be
J_\|=\sum_i\frac{\langle J_s C_i\rangle}{\langle C_i^2 \rangle} C_i .\label{jp}
\ee
$J_\|$ can be interpreted as the projection of the the spin current to the space of conserved quantities, i.e.~the conserved part of $J_s$ and, indeed, one obtains directly from Eq.~(\ref{suz}):  
\be
D_s= \frac{\beta}{N} \langle J_\|^2 \rangle . \label{dj}
\ee

As described above, the known local conservation laws $Q_n$ do not contribute to $J_s$, i.e.~$\langle J_\| Q_n \rangle=0$. $J_\|$ is a very complex non-local operator which is difficult to construct and handle analytically. For finite size systems with up to 20 sites, however, one can construct $J_\|$ numerically using the exact eigenstates of $H_0$.
As the 
$C_i$ span the space of energy diagonal operators, we just keep the energy diagonal part of $J_s$, i.e.~
\be
\langle n| J_\| | m \rangle=\delta_{E_m E_n} \langle n| J_s | m \rangle .\label{proj}
\ee

For a finite value of the perturbation $J'$ the Drude weight (\ref{dj}) is absent, as is known 
from numerical studies \cite{prelovsek:2004,heidrich-meisner:2003} which were, however, not able to investigate the regime of small $J'$ due to large finite size effects in this limit. 

In Ref.~[\onlinecite{jung.rosch:2007}] we have shown that a lower bound
for the leading order contribution to the conductivity $\s_s$ can be obtained  in the limit of small $J'$ by evaluating the correlation function $\tilde{\Gamma}$ with respect to $H_0$: 
\be
\Re\tilde{\Gamma}(\w)= \frac{1}{N} \int_0^\infty dt e^{i \w t} \langle [\dot{J_\|}(t), \dot{J_\|}(0)]\rangle_0 . \label{G2}
\ee
As $[J_\|, H_0]=0$, $\tilde{\Gamma}(\w)$ is proportional to $J'^2$.
The inequality for the spin conductivity reads
\be
 \s_s \geq \frac{\chi^2}{\tilde{\Gamma}(0)}
\ee
where $\chi=\beta \langle J_\| J_s \rangle/N=D_s$ is the generalized (spin current) susceptibility and  $\tilde\Gamma(\w)/\chi$ can be interpreted as a scattering rate of $J_\|$, see 
Ref.~[\onlinecite{jung.rosch:2007}] for details. Next we will present our analysis of the correlation function Eq.~(\ref{G2}).

\section{Numerical results} 
We investigate $\tilde{\Gamma}(\w)$ 
and the generalized susceptibility $\chi$ numerically in the $T\to \infty$ limit 
via exact diagonalization for system sizes up to $L=20$ and for various 
anisotropies $\Delta$ using periodic boundary conditions. In this high temperature limit, the spin-spin correlation length vanishes and therefore finite size effects are smallest. Results for finite $T \gg J$ (not shown) are essentially identical.

\begin{figure} \begin{center} 
\includegraphics[width=.98\linewidth,clip=]{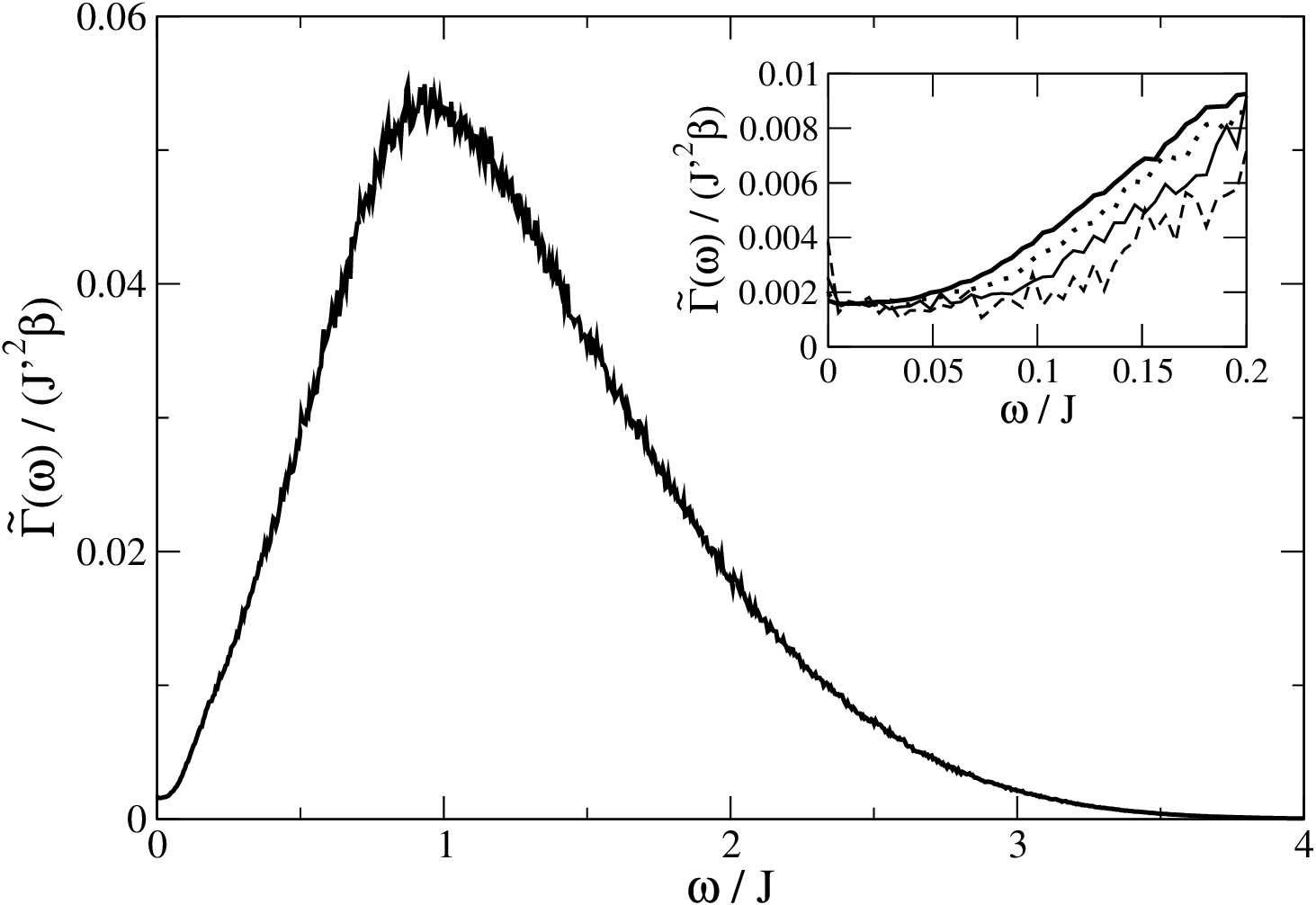} 
\end{center} \caption{\label{fig1} Leading order contribution to the spin current relaxation rate for $\Delta=0.75$ and system size $L=20$ for $T\to \infty$. Finite size effects are small and $\tilde{\Gamma}(\w)$ is finite at $\w=0$ as can be seen in more detail in the inset (thick line $L=20$, dotted line $L=18$, thin line $L=16$, dashed line $L=14$).}
\end{figure}
The results for an intermediate $\Delta=0.75$ are shown in Fig.~\ref{fig1}. 
$\tilde{\Gamma}(\w)$ drops rapidly for small frequencies but saturates at a finite value. This saturation value $\lim_{\w \to 0} \tilde{\Gamma}(\w)$ is almost independent of system size (see inset). This indicates that finite size effects are small despite the fact that $J_\|$ is expected to be a non-local operator. We therefore conclude that for small $J'$
\be
\sigma_s\geq \frac{c(T)}{T J'^2}.
\ee
in the thermodynamic limit. This is the main result of this paper: the spin-conductivity of a slightly perturbed XXZ Heisenberg chain is very large, despite the fact, that the spin current is not protected by any local conservation law. For $\Delta=0.75$ we obtain for example $c(T\to \infty)=0.92 J^3$. For any finite temperature we expect that the same result holds: 
in the limit of small $J'$  the spin conductivity is proportional to $1/J'^2$.

\begin{figure} \begin{center} 
\includegraphics[width=.98\linewidth,clip=]{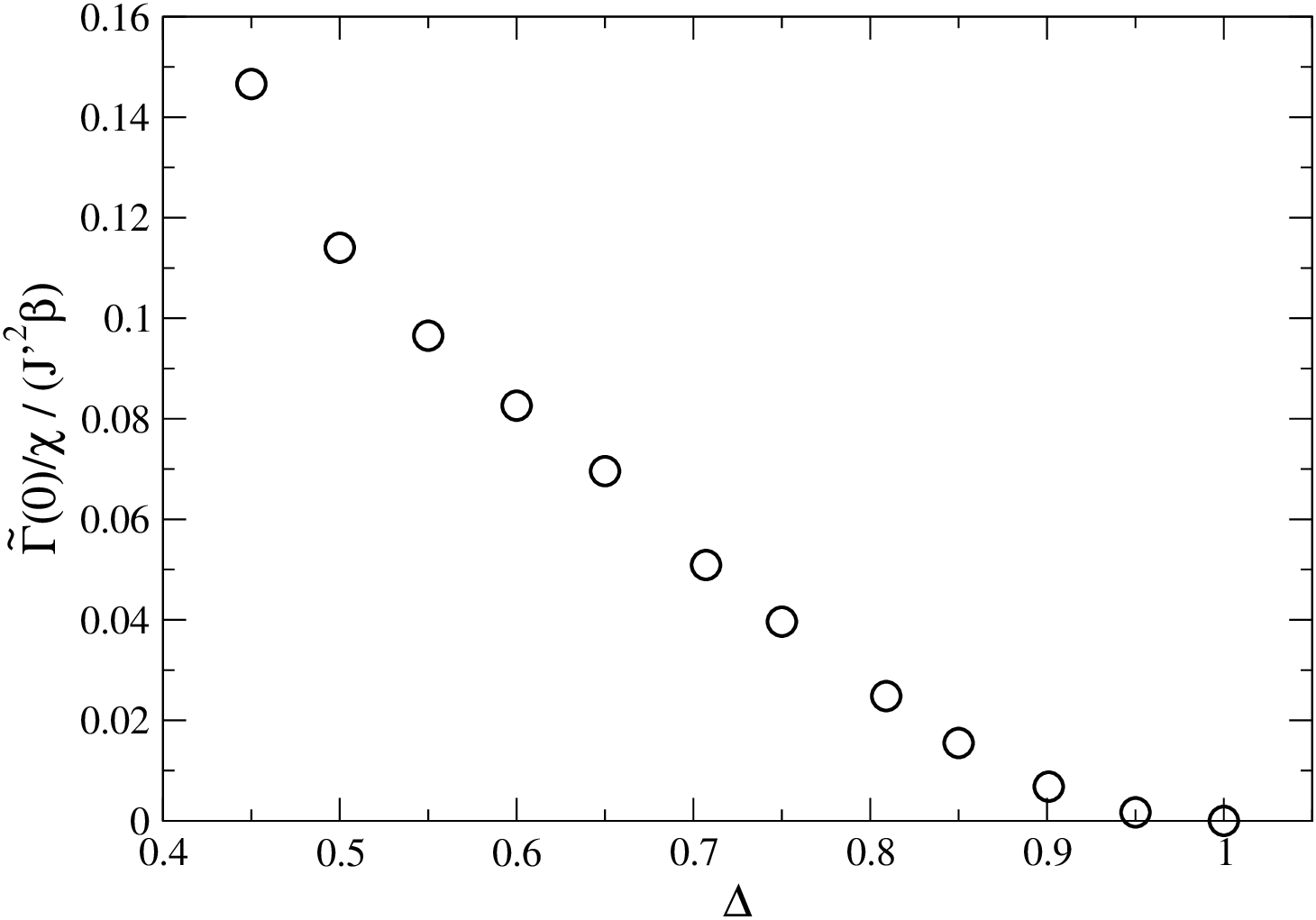} 
\end{center} \caption{\label{fig2} Scattering rate  $\tilde{\Gamma}(0)/\chi$ as a function of the anisotropy parameter $\Delta$ for  $L=18$, $T=\infty$. For the isotropic system, $\Delta=1$, 
$\tilde{\Gamma}(0)$  is zero, see text. The errors are comparable to the size of the symbols and are discussed in more detail in Appendix \ref{appA}.}
\end{figure}
In Fig.~\ref{fig2} the behavior of the scattering rate $\tilde\Gamma/\chi$ as a function of $\Delta$ is shown. Interestingly, the scattering rate seems to vanish in the isotropic limit $\Delta \to 1$, $\tilde\Gamma \propto J'^2 (1-\Delta)^2$. We have  previously  \cite{jung.rosch:2006} observed the same effect for the scattering rate of the heat current, which turns out to be proportional to $1/J'^4$ at the isotropic point. The reason for this unexpected result is that for the isotropic case one can construct an operator $Q_3'=Q_3+J' \Delta Q_3$ such that the commutator $[Q'_3,H_0+H_1]$ is of order $J'^2$ rather than linear in $J'$. As a consequence, the decay rate of the heat current at the isotropic point is proportional to $J'^4$. Very likely, the same mechanism applies to $J_\|$, too. A subtle and controversial issue \cite{zotos:1999,heidrich-meisner:2003,klumper:2005} is the value of the Drude weight, $D_s=\chi$, for $\Delta=1$. Both from numerics and from Bethe ansatz, there is evidence pointing either to a finite\cite{heidrich-meisner:2003,klumper:2005} or vanishing \cite{zotos:1999,klumper:2005} Drude weight in the thermodynamic limit. If the Drude weight vanishes for $\Delta=1$, our results are only of relevance for $\Delta<1$.

In appendix \ref{appA} we discuss  a further effect: the Drude weight $D_s$  appears to be a discontinuous function of $\Delta$ as for special values of the anisotropies $\Delta=\cos(\pi/n)$, $n=3,4,5...$, one obtains different values for $D_s$ compared to anisotropies slightly away from these points. For the scattering rate $\tilde\Gamma/\chi$ these effects are much smaller and possibly absent in the thermodynamic limit.

\section{Non-locality of $J_\|$}\label{appB}

As stressed in the introduction, the spin current $J_s$ is orthogonal to all know local conservation laws $Q_n$ of the XXZ Heisenberg chain. This suggests that $J_\|$, the conserved part of $J_s$, is a non-local operator which cannot be written in the form of Eq.~(\ref{local}).
To quantify this statement, we expand the numerically constructed $J_\|$ in local operators $A_{ni}$  which contain products of spin-operators on $n$ adjacent sites,
\bea
J_\|=\sum a_{ni} \frac{A_{ni}}{\langle A_{ni}^2 \rangle^{1/2}} , \label{expand}
\eea
where the $A_{ni}$ define a complete orthogonal basis in the space of operators, $\langle A_{ni} A_{mj} \rangle=0$ for $n\neq m$ or $i\neq j$. The $A_{ni}$ are written as sums of products of spin-operators, where each product contains spins on $n$ adjacent sites. Here we use -- as above -- the  ($T=\infty$) expectation value as the scalar product in the space of operators. In Eq.~(\ref{expand}) obviously only translationally invariant hermitian operators contribute which also conserve $S_z$. For $n=1$ there is just one such operator, $A_{11}=\sum_i S^z_i$, for $n=2$ one finds $3$ such terms 
$A_{21}=\sum_i S^z_i S^z_{i+1}$, $A_{22}=\sum_i S^+_i S^-_{i+1}+h.c.$, $A_{23}=i \sum_i( S^+_i S^-_{i+1}-h.c.)$. The 10 operators of range $3$, $A_{3i}$, contain both products of two spin operators, e.g.~$\sum_i S^z_i S^z_{i+2}$ and products of three spin-operators, e.g.~$\sum_i S^z_i S^z_{i+1} S^z_{i+2}$.

\begin{figure} \begin{center} 
\includegraphics[width=.98\linewidth,clip=]{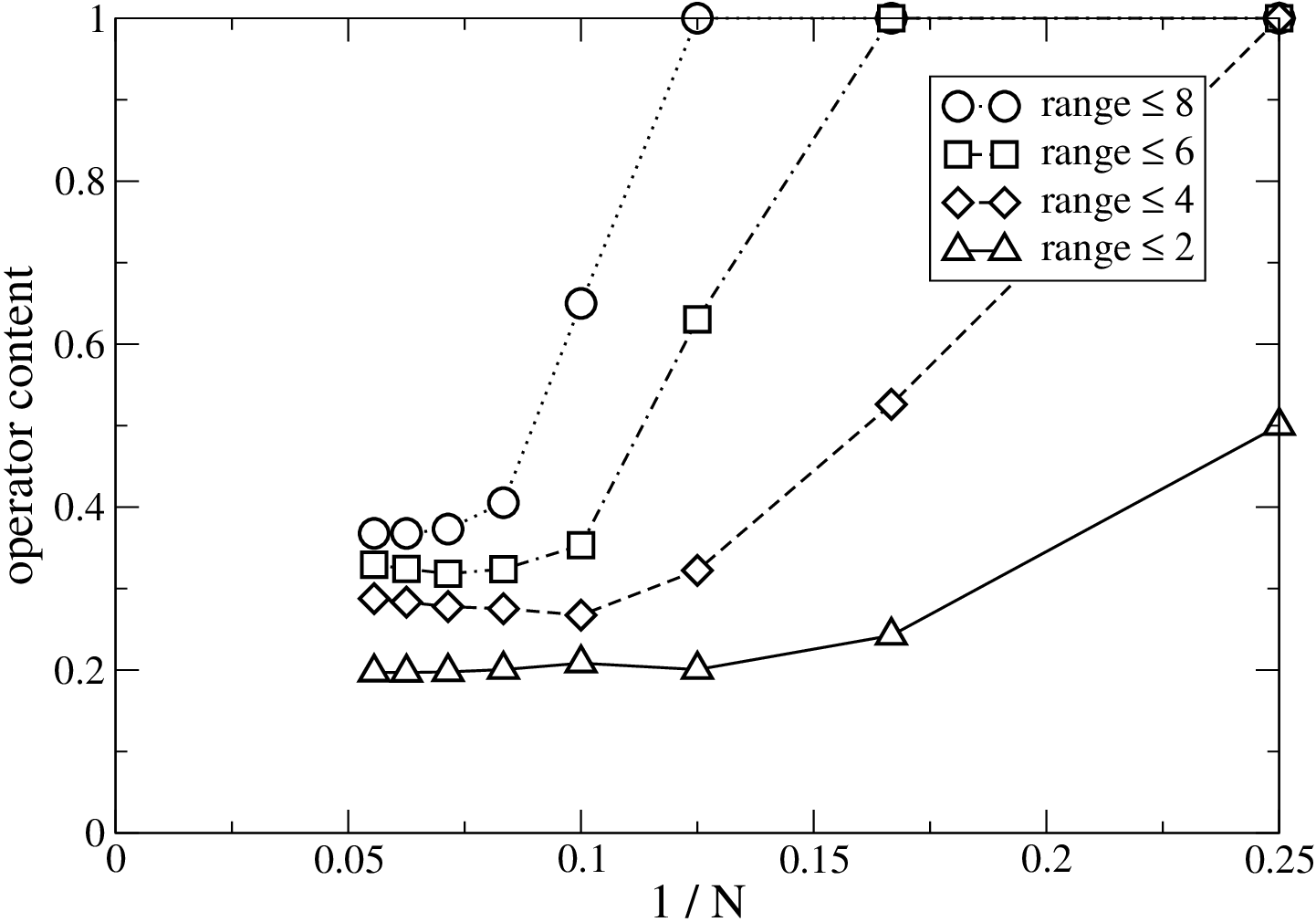} 
\end{center} \caption{\label{range} Relative weight $\sum_{m=1}^n c_m$, see Eq.~(\ref{cn}), of local operators with range up to $n$ ($n=2,4,6,8$) contributing to  $J_\|$  as a function of inverse system size from $N=4$ to $N=18$. }
\end{figure}
\begin{figure} \begin{center} 
\includegraphics[width=.98\linewidth,clip=]{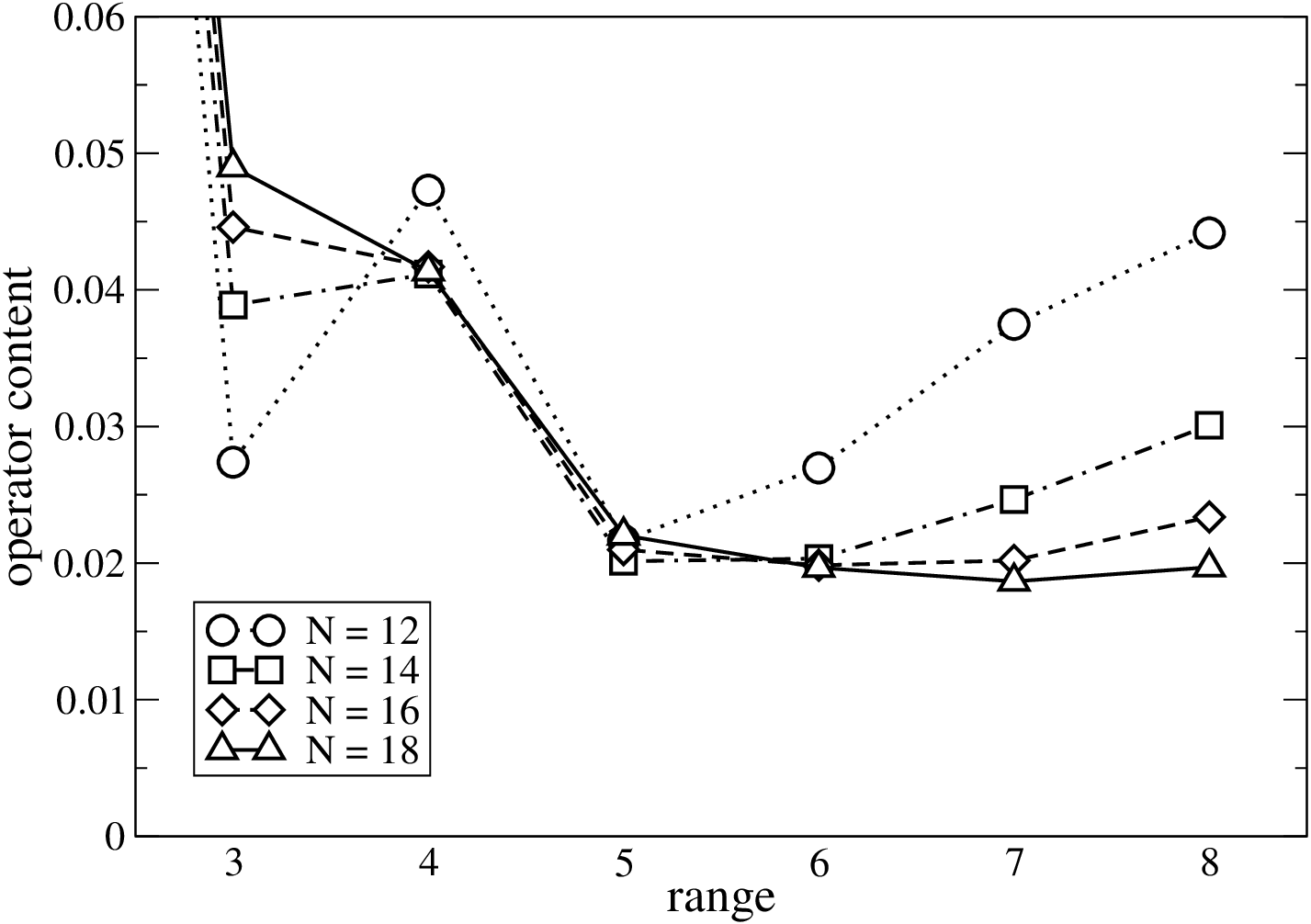} 
\end{center} \caption{\label{weights} Relative weight $c_n$ of local operators of range $n$ ($n=3,\dots 8$) contributing to  $J_\|$ for different system sizes. Note that in the thermodynamic limit most of the weight is carried by operators involving more than 8 consecutive sites 
(see Fig.~\ref{range}). }
\end{figure}
The ratio
\bea
c_n = \frac{\sum_{i} |a_{ni}|^2}{\sum_{i,m} |a_{mi}|^2} \label{cn} 
\eea
shown in Figs.~\ref{range} and \ref{weights} describes which fraction of the operator $J_\|$ can be expressed in terms of operators of range  $n$. For example, if one determines the $c_n$ for $H$ one obtains $c_2=J^2/\left( J^2+J'^2\right)$ and $c_3=J'^2/\left( J^2+J'^2\right)$. By construction one gets $\sum_{n=0}^N c_n=1$ for a system with $N$ sites.

What types of behavior can be expected for $c_n$? 
First, one has to investigate whether $c_n$ is finite or zero in the thermodynamic limit $N\to \infty$. For example, for the square of a translationally invariant local operator (e.g. $H_0^2$), one finds that $c_n$  drops proportionally to $1/N$, such that  $\lim_{N\to \infty} c_n =0$ for all $n>0$. Even if $\lim_{N\to \infty} c_n$ is finite for each $n$, one can ask how rapidly $\lim_{N\to \infty} c_n$ drops for $n\to \infty$ and whether $\sum_{n=0}^\infty \lim_{N\to \infty} c_n$ equals $1$ or is smaller. 

As shown in Fig.~\ref{range}, the $c_n$ converge to finite values for $N \to \infty$. For $n=2$, this is a necessary consequence of the fact that the spin current is a range 2 operator and that the Drude weight of the spin current is finite. As the latter is proportional to $\langle J_s J_\| \rangle^2$ this implies that $J_\|$ has a finite overlap with a range 2 operator in the thermodynamic limit.

A qualitative result of Fig.~\ref{range}  is, however, that even operators up to range 8 have less than 40\% of the total weight of $J_\|$ (but $c_8\approx 0.02$ is already very small). As $\sum_{n=1}^\infty c_n=1$, the $c_n$ have to drop faster than $1/n$ for large $n$ in the thermodynamic limit. Fig.~\ref{weights} shows that the $c_n$ decay extremely slowly with $n$. In this sense $J_\|$ appears to be a rather nonlocal operator
but we cannot decide from our numerics whether $\sum_{n=0}^\infty \lim_{N\to \infty} c_n=1$ or smaller.

\section{Conclusion} 

In this paper, we have shown that the spin-conductivity of a one-dimensional anisotropic spin-chain is strongly enhanced close to the integrable point. It diverges (at least) as $1/J'^2$ for $J' \to 0$. This is the expected behavior for a situation where a local conservation law prohibits the decay of the current at the integrable point. However, as emphasized by Zotos, Naef and Prelovsek \cite{zotos:1997}, the spin-current is orthogonal to all known local conservation laws of the XXZ chain. 

There are two possible interpretations of this result. First, the conserved part $J_\|$ of the spin current could nevertheless be `sufficiently' local to define a slow hydrodynamic mode. Second, the theoretical prejudice, that only local conservation laws (i.e.~those associated with a continuity equation) lead to slow modes, may be wrong. In this respect, the results of section \ref{appB}, where this question is investigated, are ambiguous. On the one hand, we could prove that $J_\|$ is a highly non-local operator involving products of operators acting on widely separated sites. On the other hand, the relative weight of range-$n$ operators, $c_n$, is finite in the thermodynamic limit. 

In this paper we have shown that the transport properties of simple one-dimensional problems depend quantitatively and qualitatively on `exotic' and rather complex conserved quantities. For the future, it would be interesting to gain a more analytic understanding of these conservation laws.
\acknowledgements

We thank A. Kl\"umper, J. Sirker and X.~Zotos for inspiring discussions. 
Part of this work was supported by the German Israeli 
Foundation and also the
Deutsche Forschungsgemeinschaft through SFB 608.

\appendix

\section{Spin-conductivity close to and at $\Delta = \cos \pi/n$}\label{appA}
\begin{figure} \begin{center} 
\includegraphics[width=.98\linewidth,clip=]{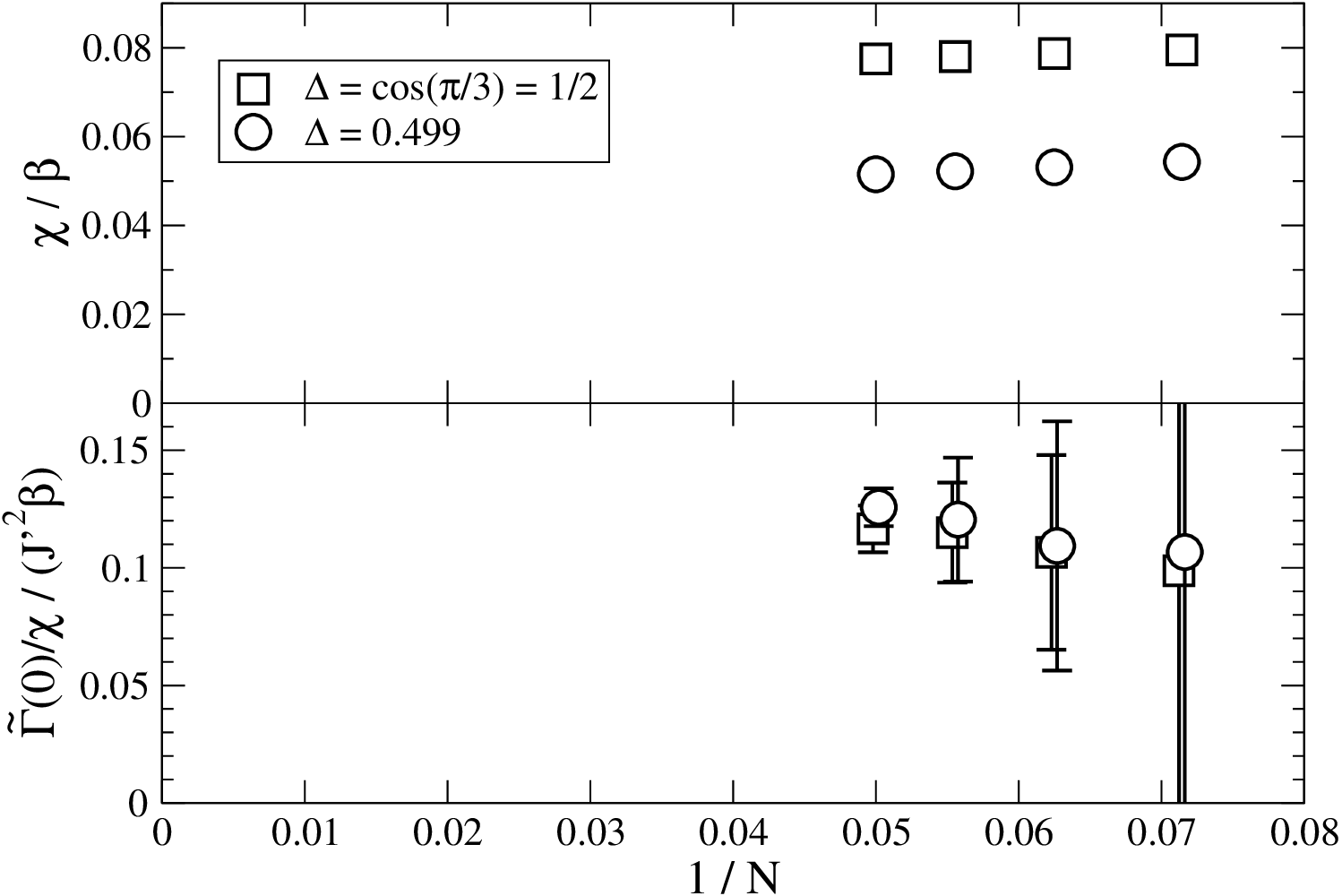} 
\end{center} \caption{\label{cnc} Drude weight $D=\chi$ (upper panel) and scattering rate $\tilde\Gamma(0)/\chi$ (lower panel) for the anisotropy  $\Delta=\cos(\pi/3)=1/2$  and a nearby value $\Delta=0.499$ as a function of inverse system size $1/N$. The error bars represent the uncertainty inherent in the fitting procedure.}
\end{figure}

In this appendix we discuss the behavior of the spin conductivity for anisotropies $\Delta=\cos(\pi/\nu)$. At these special points it is known that there are further symmetries which, for example, simplify the Bethe ansatz equations considerably\cite{deguchi:2006,sirker.bortz:2006}. Interestingly, at these special points thermodynamic quantities show unexpected logarithmic corrections  \cite{sirker.bortz:2006}.

In Ref.~[\onlinecite{naef.zotos:1998}], Naef and Zotos found numerically that the Drude weight at these special points differs for finite systems significantly from the values obtained for slightly different anisotropies. They concluded, however, that these differences vanish in the thermodynamic limit. While we have reproduced the numerical results of Naef and Zotos, we interpret our results differently. In Fig.~\ref{cnc} we show the Drude weight of the integrable model, $D_s=\chi$, and the scattering rate, $\tilde{\Gamma}(0)/\chi$ , as a function of $1/N$ both for $\Delta=\cos \pi/3=1/2$ and $\Delta=0.499$. Apparently, different values are obtained for $N\to \infty$ for $\chi$ (subtle  logarithmic finite size effects can possibly invalidate this analysis)  while the effect for the scattering rate $\tilde{\Gamma}(0)/\chi$ is much smaller (and possibly absent).



\end{document}